# 0.52 V-mm ITO-based Mach-Zehnder Modulator in Silicon Photonics

Rubab Amin,[1] Rishi Maiti,[1] Caitlin Carfano,[1] Zhizhen Ma,[1] Mohammad H. Tahersima,[1] Yigal Lilach,[2] Dilan Ratnayake,[2] Hamed Dalir,[3] and Volker J. Sorger[1,a]

[1]Department of Electrical and Computer Engineering, George Washington University, 800 22nd St., Washington, District of Columbia 20052, USA

[2]Nanofabrication and Imaging Center, George Washington University, 800 22nd St., Washington, District of Columbia 20052, USA

[3]Omega Optics, Inc. 8500 Shoal Creek Blvd., Bldg. 4, Suite 200, Austin, Texas 78757, USA

Electro-optic modulators transform electronic signals into the optical domain and are critical components in modern telecommunication networks, RF photonics, and emerging applications in quantum photonics and beam steering. All these applications require integrated and voltage-efficient modulator solutions with compact formfactors that are seamlessly integratable with Silicon photonics platforms and feature near-CMOS material processing synergies. However, existing integrated modulators are challenged to meet these requirements. Conversely, emerging electro-optic materials heterogeneously integrated with Si photonics open a new avenue for device engineering. Indium tin oxide (ITO) is one such compelling material for heterogeneous integration in Si exhibiting formidable electro-optic effect characterized by unity order index at telecommunication frequencies. Here we overcome these limitations and demonstrate a monolithically integrated ITO electro-optic modulator based on a Mach Zehnder interferometer (MZI) featuring a high-performance half-wave voltage and active device length product, $V_\pi L$ = 0.52 V·mm. We show, how that the unity-strong index change enables a 30 micrometer-short π-phase shifter operating ITO in the index-dominated region away from the epsilon-bear-zero ENZ point. This device experimentally confirms electrical phase shifting in ITO enabling its use in multifaceted applications including dense on-chip communication networks, nonlinearity for activation functions in photonic neural networks, and phased array applications for LiDAR.

## I. INTRODUCTION

Current communication systems utilize coherent detectors and advanced modulation formats that encode data onto both the amplitude and phase of the light signals. Optical modulation arises from the controlled variation in the complex refractive index of the active (modulated) material, where phase modulators utilize the change in the real part of the optical index [1, 2]. Electro-optic modulation occurs from the electrical control of the optical index change and can be divided into two major groups, i.e. phase and absorption modulation, based on the corresponding controlled change in the real or imaginary part of the active material's index [3]. In phase modulation a change in the refractive index, $n$, and the Kramers-Kronig (K-K) relations dictate the change in the material absorption, corresponding to the extinction coefficient, $\kappa$ [1-3]. Phase modulators require some form of interferometric scheme to evoke an amplitude change, such as the Mach-Zehnder configuration, interestingly already over a century old [4, 5], whereas absorption modulators realize this in a (simpler) linear waveguide geometry. Mach-Zehnder interferometers (MZIs) and modulators (MZMs) have been applied in both free space and integrated optics domains for phase change applications and sensing, whilst cavity feedback schemes have also been investigated extensively for phase modulation, however trading in spectral bandwidth [6-8].

---

[a] Author to whom correspondence should be addressed. Electronic mail: sorger@gwu.edu.

The performance of a MZM depends inherently on the underlying physical modulation mechanisms, such as electric field based Pockels effect, Kerr effect, quantum confined stark effect (QCSE), and free carrier modulation as found in Silicon (Si) and transparent conductive oxides (TCOs) to include Indium Tin Oxide (ITO), which is a highly doped degenerate semiconductor that the index tuning was well studied by carrier-dependent Drude model (i.e. current-driven modulators) [6-14]. However, unlike Si, the carrier concentration of ITO can be (a) higher and (b) more dramatically tuned [1-3,9-12]; the stronger modulation-per-charge carrier change of ITO (compared to Si) originates from the larger bandgap, and consequently, lower permittivity of ITO [1,11], and has experimentally shown index changes of about unity-order under electrical gating [9,10,15]. Previously we showed that ITO can be used as an electro-optic (EO) case in the $n$–dominant regime, or for the electro-absorptive (EA) cases in the $\kappa$–dominant regime depending on the bias (carrier) condition and relative position to the epsilon-near-zero (ENZ) point [1-3,11]. Here, we experimentally demonstrate a heterogeneous integrated ITO-based MZM in Si photonics showing a $V_\pi L$ of 0.52 V·mm at $\lambda$ = 1550 nm, operating at an optimized index-to-loss ratio, i.e. $(\delta\lambda/\delta\kappa)|_{min}$, away from the ENZ region.

In the following discussion the nomenclature regarding the ON-OFF states of operation for the modulator relates to the light transmission (ON) vs. low-transmission (OFF) characteristics through the device, rather than the applied voltage bias.

**II. RECENT ADVANCES IN RELEVANT FIELD**

Since phase modulators require interferometric schemes (e.g. ring resonators, MZIs, etc.) they do inherently suffer from an extended footprint compared to absorption-based modulators. In MZMs, the product of the half-wave voltage times the active modulator length, $V_\pi L$, is a figure of merit (FOM), since they exhibit a tradeoff between obtaining π-phase shift with increases device length or voltage. While MZMs based on lithium niobate (LiNbO$_3$) are commercially available, their $V_\pi L$ is rather high due to the weak Pockels effect, whereas improved performance is obtained with the QCSE in III-V [26,28,30] or emerging materials such as polymers [24,27,32-34] and enhanced light-confinement for improving optical and RF mode overlap ($\Gamma$) with the active material [32-34] (Table 1). This work focuses on an experimentally demonstrated ITO based MZM device and the achieved FOM in our results seems promising despite being the first of its kind.

**Table 1: Figure of merit (FOM) comparison for Mach Zehnder devices with different active modulation materials and waveguide structures in recent years.**

| Structure/Material | $V_\pi L$ (V.mm) | Ref. |
|---|---|---|
| Thin plate LiNbO$_3$ | 90 | [16,17] |
| Domain inverted push-pull LiNbO$_3$ | 90 | [18] |
| Ridge LiNbO$_3$ | 80 | [19] |
| Metal-oxide-semiconductor (MOS) - Si | 80 | [20] |
| LiNbO$_3$ physical limit | 36 | [21] |
| Doping optimized Si | <20 | [22] |
| Si pin | 13 | [23] |
| Silicon-organic hybrid (SOH) | 9 | [24] |
| Si Lateral-pn | 8.5 | [25] |
| III-V Multiple Quantum Wells (MQW) | 4.6 | [26] |
| SOH | 3.8 | [27] |
| GaAs/AlGaAs | 2.1 | [28] |
| Hybrid Si MQW | 2 | [29] |
| InGaAlAs/InAlAs MQW | 0.6 | [30] |
| ITO MOS | 0.52 | This work |
| Si p$^+$-i-n$^+$ | 0.36 | [31] |
| EO Polymer Plasmonic | 0.07 | [32] |
| Liquid crystals with SOH slot/ all-plasmonic polymer | 0.06 | [33,34] |

**III. DEVICE DESIGN AND FABRICATION**

The inherent loss imbalance between the arms due to fabrication imperfections is a challenge in MZM schemes. This loss imbalance limits to achieve higher extinction ratio (ER) defined as the ratio between maximum and minimum output power. It originates from the complex part of the optical phase leads to degraded optical signal fidelity (i.e. phase errors at the output and alter frequency chirps) [35]. Intrinsically, phase shifting impacts both the real and imaginary parts (i.e. K-K relations), thus the loss imbalance of an MZM can alter during MZM operation between the ON/OFF states. One can improve the arm loss imbalance by tuning the MZM arm losses statically. However, a design criteria for using an inherently lossy material such as ITO, in an interferometric scheme similar to the MZ configuration to achieve satisfactory modulation depth, is to match the amplitude of the optical signal (i.e. loss) in both arms of the MZ; the interference from both arms at the output terminal converge referred hereafter to as balancing. The passive MZI is built on a silicon-on-insulator (SOI) substrate with the same waveguide lengths in both the arms, where subsequent process steps towards the active device include depositing ITO on top of a portion of one arm separated by an oxide layer to facilitate gating (Fig. 1). A symmetrical passive MZI (i.e. same length for both arms and 50/50 Y-splitters on both sides) is chosen so that the interference pattern at the output can distinguishably confer modulation effects from our active ITO device. The MZM system can be simplified by combining the field loss and phase shifts through each arm. The output field is then a simple function of the input field given by

$$\tilde{E}_{out} = \tilde{E}_{in}\left(a_1 e^{-i\phi_1} + a_2 e^{-i\phi_2}\right) \qquad (1)$$

where $a_1$ and $a_2$ are the field gain (loss) in each arm of the MZI and $\phi_1$ and $\phi_2$ are the phase induced through each of the arms (active/modulated and un-modulated), respectively. The input and output fields are denoted by phasor quantities, i.e. $\tilde{E} = E e^{i\omega t}$, assuming there is no gain in the system, $(a_1 + a_2)^2 \leq 1$. The time-dependent transfer function for the light intensity (or power) using the slowly-varying envelope approximation (modulation frequency << optical carrier frequency) is expressed as

$$T = \left|\frac{E_{out}}{E_{in}}\right|^2 = a_1^2 + a_2^2 + 2a_1 a_2 \cos\Delta\phi \qquad (2)$$

where $\Delta\phi$ is the phase difference between the arms, $\Delta\phi = \phi_1 - \phi_2$. To maximize the obtainable *ER*, i.e. ensuring minimal zeros in the OFF state, the field losses in both arms need to be matched i.e. $a_1 = a_2$. Deviations from this ideal case are typically attributed to imperfect 50/50 Y-couplers [36, 37]. However, it is critical to emphasize that deviation from $a_1 = a_2$ can be a result from differences in the losses anywhere in the MZM configuration including possible fabrication imperfections. In contrast, higher index change materials (e.g. ITO) do accompany significant loss as a byproduct of modulation and as such both the states of operation need to be accounted for in design considerations.

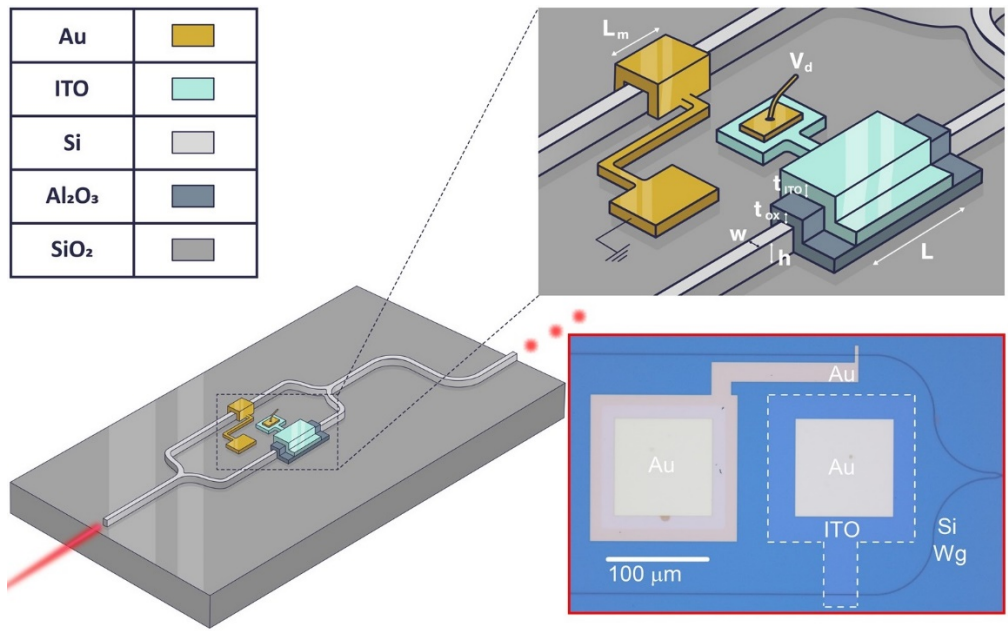

FIG. 1 Schematic of the ITO-based Mach Zehnder Interferometer (MZI) showcasing the active gating region and contacts. The metal (Au) on the un-modulated arm serves as a contact for the capacitor while adding necessary loss for balancing the device. An optical microscope image of the fabricated device showing the active modulation region and contacts is shown in the inset, the dashed outline marks the patterned and deposited ITO thin film. Relevant parameters are: Length of the metal contact on Si, $L_m$ = 3.7 µm; active device length, $L$ = 32 µm; thickness of the deposited ITO thin film, $t_{ITO}$ = 10 nm; thickness of the Al$_2$O$_3$ gate oxide, $t_{ox}$ = 10 nm; waveguide height, $h$ = 220 nm and width, $w$ = 500 nm. Image not drawn to scale.

The extinction ratio (ER) is the ratio of the transmission between the ON- ($T_{max}$) and OFF state ($T_{min}$), i.e. static ER since it is measured by varying a DC phase bias to one of the arms to find the absolute maximum and minimum transmission. This is necessary since the dynamic ER may be reduced when operating at high frequencies due to limited phase swings or pulse shaping from the finite bandwidth of the electrodes. This upper bound can be referred to as the maximum extinction ratio, $ER_{max}$. Defining $\gamma = a_2/a_1$ as the field loss imbalance in the ON-state and, similarly $\gamma' = a'_2/a_1$ for the OFF state, where $a'_2$ denotes the OFF-state field loss since the change in field loss is not negligible for ITO resulting from the K-K relations, the maximum extinction ratio can be expressed as,

$$ER_{max} = \left(\frac{1+\gamma}{1-\gamma'}\right)^2 \tag{3}$$

Here, the field losses can be approximated as, $a = e^{-\alpha L}$ where $L$ is the device length and the absorption due to the altered ITO material is $\alpha = 2\pi\kappa_{eff}/\lambda$; where $\lambda$ is the operating wavelength, and $\kappa_{eff}$ is the imaginary part of the effective index. With the aim to aim for $ER_{max}$ and adjusting for both states of operation, we calculate the desired length of the metal contact, $L_m$ on the un-modulated arm of the MZM, and chose to deposit metal (Au) on the other (un-modulated) arm of the MZM (Fig. 1) for two reasons: (a) it acts as our bottom electrode in the metal-oxide-semiconductor (MOS) stack (i.e. the Si waveguide is lightly doped); and (b) imposes necessary loss on the un-modulated arm to facilitate modulation depth, i.e. balancing the loss in both the arms.

Regarding ITO material modeling, the Drude model characterizes the ITO material response well within the near infra-red region, and our deposited ITO thin films exhibit similar optical indices (real and imaginary parts) as obtained from Drude model fits in previous works [1-3,11]. Spectroscopic ellipsometry of deposited thin films show a decreasing real part, $n$ and an increasing imaginary part, $\kappa$ of the material index near our operating wavelength, $\lambda$ = 1550 nm (Fig. 2a), which is expected from the K-K characteristics [1-3]. When ITO is packaged as one electrode of an electrical capacitor, applying bias voltage can put the capacitor into the three known states of accumulation, depletion, or inversion, thus changing the carrier concentration. The optical property of the active material therefore changes significantly, resulting in strong optical modulation. In praxis, a 1/e decay length of about 5 nm has been measured before [38], and modulation effects have been experimentally verified over 1/e$^2$ (~10 nm) thick films from the interface of the oxide and ITO [9]. In order to extract relevant parameters including the effective indices (real and imaginary parts, $n_{eff}$ and $\kappa_{eff}$) and confinement factors, $\Gamma$, we perform FEM eigenmode analysis for our structure. The first order transverse magnetic (TM)-like mode is selected following the TM-optimized grating couplers in the fabricated device and the mode profiles indicate an increase in the light confinement with modulation by 41% which is aligned with results from our previous work as we operate away from the ENZ point in the *n*-dominant region (Fig. 2b) [1, 2, 11].

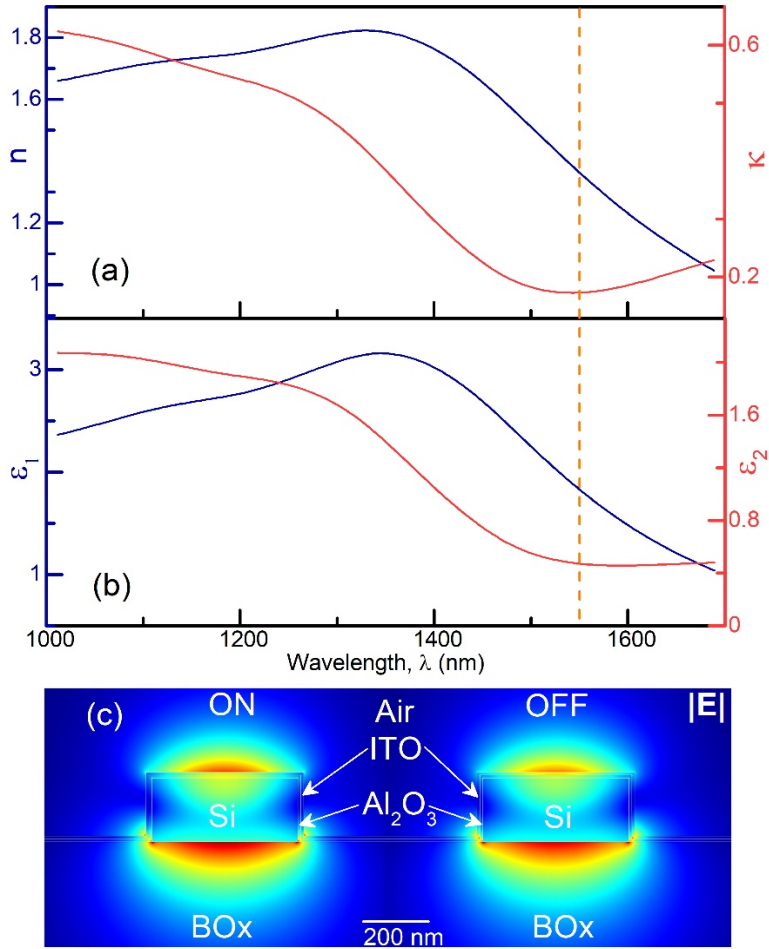

FIG. 2 Material and modal characteristics. (a) Material refractive index (real and imaginary parts, *n* and *κ*) and (b) Permittivity (real and imaginary parts, $\varepsilon_1$ and $\varepsilon_2$) vs. wavelength, $\lambda$ (nm) obtained from spectroscopic ellipsometry of our ion beam deposited ITO. The dashed orange line represents the operating wavelength at $\lambda = 1550$ nm. (c) Mode profiles (|E|-field) for our structure along the active device corresponding to states of operation. The decreased leakage in air in the OFF state indicates increased confinement in the ITO layer compared to the ON state, suggesting an increase in LMI.

From our Eigenmode simulations (Fig. 2c) and relevant analytical expressions, we find the loss needed to balance the MZI. Since the modulation efficiency (ER/$V_{pp}$) is improved for better electrostatics, we use a relatively high-*k* dielectric, $Al_2O_3$, for the gate oxide of 10 nm using atomic layer deposition (ALD), followed by 10 nm of ITO on the device region (ion beam deposition (IBD)). The latter has synergies for processing ITO as this process yields density films that are pinhole-free and highly uniform, and allows for a room temperature process, which does not anneal ITO (i.e. no activation of tin carriers as to facilitate electrostatic EO tuning). Incidentally, IBD technologies are advantageous for nanophotonic device fabrication due to their precise controllability of material properties such as microstructure, non-stoichiometry, morphology, crystallinity, etc. [39, 40]. ~~50 nm of Au was used for the contacts as it has reasonably low ohmic loss at near IR wavelengths. An additional 3 nm adhesion layer of Ti was used in the contacts.~~ The fabricated device is shown in the optical microscope image in Fig. 3(a); white lines denote the patterned 10 nm ITO thin film designating the active device region on top of the corresponding Si waveguide (actively modulated arm of the MZM). The bottom Au contact for the MOS stack is on the other MZM arm (Fig. 3a). As the ALD does not allow patterned processing due to chamber contamination concerns, an etch step was required on top of the Si contact to remove the oxide over it for electrical probing. We used a rather slow wet etch process for $Al_2O_3$ using an MF319 solution in the contact pad area (See supplementary information). The etched oxide opening on top of the metal pad can be noticed from the color contrast difference (Fig. 3a). Finally, two contact pads side by side are deposited on both contact areas to facilitate biasing (marked with Au in Fig. 3a). The ITO contact is used to administer the voltage while the bottom Si contact is grounded. TM-optimized grating couplers are used to couple the light from (to) the fiber into (out of) the MZM.

**IV. RESULTS AND DISCUSSION**

Experimental results show a modulation depth (i.e. ER) of ~2.1 dB for an phase shifter length of only 32 μm (Fig. 3a). The voltage needed for π-phase shifts at the output is about 16 V (Fig. 3c) gives a corresponding $V_\pi L$ of just 0.52 V·mm. The output intensity of the MZ configuration is governed by [41]

$$I \propto |E|^2 \propto \cos^2\left(\frac{\Delta\beta L}{2}\right) \qquad (4)$$

where, the output intensity is normalized such that the peak transmission factor is 1 for ideal power transmission, $\Delta\beta$ is the induced change in the propagation constant, $\beta$ between both arms during modulation. Fitting the experimental data to Eqn (4) we extract the half wave voltage, $V_\pi$. The change in the effective index of the waveguide, $\Delta n_{eff} = 0.020$, for ON/OFF modulation is estimated using the applied voltage, $V_d$ and the material index change in the ITO, $\Delta n_{ITO}$ from Figure 3d,e, which is closely match the value obtained from the FEM analysis (~0.023). The effective index change with modulation can be found by a linear approximation with applied voltage as $\partial n_{eff}/\partial V_d \sim 1.407 \times 10^{-3}$ V$^{-1}$. Both the effective index change, $\Delta n_{eff}$ and material index change in ITO, $\Delta n_{ITO}$ exhibit monotonic increase with bias. This is expected as we operate in the *n*-dominant region of the ITO material far from any ENZ effects. Note, the change in both the indices correspond to a decrease in the corresponding indices as modulation assimilates to blue-shifts in device resonance, however is hardly resolvable in our single pass MZ configuration (Fig. 3b), but is well-known from ring resonators, photonic crystal cavities or any Fabry-Pérot cavity. The weak dispersion of $\partial n_{eff}/\partial \lambda \sim 1.12 \times 10^{-4}$ nm$^{-1}$ in our modal structure contributes to the undistinguishable resonance shift in the transmission spectra. The modal dispersion is calculated from Eigenmode analysis. The material index change in ITO, $\Delta n_{ITO}$ shows near unity order index change as demonstrated and denotes the potential of this emerging material (Fig. 3e), which is significantly higher as compared to its Si counterpart while both (ITO and Si) operate with the free carrier modulation mechanism. This improvement of ITO can be attributed to: (a) 2-3 orders higher carrier density, and (b) the higher bandgap, which consequently leads to a lower refractive index [1,2]. If the change of the carrier concentration $\delta N_c$ (e.g. due to an applied voltage bias) causes a change in the relative permittivity (dielectric constant) $\delta\varepsilon$, the corresponding change in the refractive index can be written as $\delta n = \delta\varepsilon^{1/2} \sim \delta\varepsilon/2\varepsilon^{1/2}$; hence, the refractive index change is greatly enhanced when the permittivity, $\varepsilon$ is small. The noticeable change from the monotonic effective indices trend originates from modal confinement increasing with bias, i.e. as we tune towards the ENZ region the confinement increases without actually biasing it to the ENZ region [1, 2, 11].

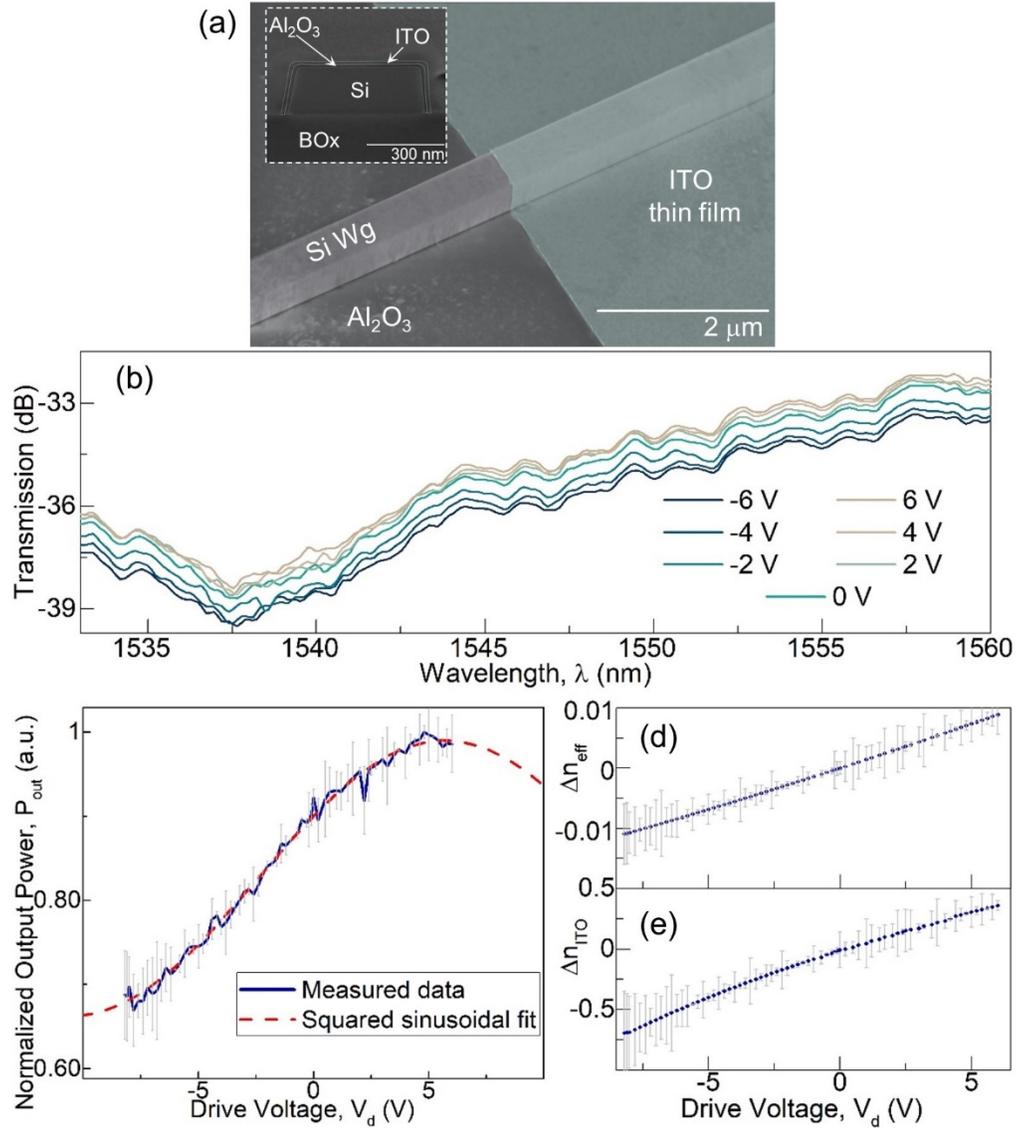

FIG. 3 Fabricated device measurements. (a) Scanning electron microscope (SEM) image of the fabricated device showing the deposited ITO thin film over the Si waveguide in the active modulation region, atomic layer deposited (ALD) $Al_2O_3$ is everywhere on top of the waveguide and underneath the ITO film. A focused ion beam (FIB) SEM cross-section of the active device revealing the oxide/ITO capacitive stack on top of the Si waveguide is shown as an inset; (b) The transmission spectra (in dB) of the device for varying drive voltages vs. wavelength, $\lambda$ (nm); (c) Output optical power, $P_{out}$ ($\mu$W) vs. drive voltage, $V_d$ (Volts); the dashed line represents a squared sinusoidal fit ($cos^2(arg)$) suggested by the underlying physics of the MZ scheme; (d) Extracted effective index change, $\Delta n_{eff}$; and (e) ITO material index change, $\Delta n_{ITO}$ with applied bias, $V_d$ (Volts) corresponding to modulation effects; the ITO index observes near unity order change with applied bias.

The simple biasing scheme used here to repurpose the Si waveguide as a bottom contact in the MOS-stack, however, severely limits modulation speed of this device due to high electrical resistance; the epi-Si layer of the SOI substrate is only lightly doped thus $R_{Si}$~600 MΩ. The contact and sheet resistance of the ITO film is ~220 Ω and 63 Ω/□, respectively (See supplementary information). The resistivity and mobility of the ITO film is measured to be 6.36 × 10$^{-4}$ Ω-cm and 42.6 cm$^2$/V-s, respectively (See supplementary information). The device capacitance is ~170 fF and hence speed is only ~1.5 kHz. Because of the design decision to employ the Si-contact on the un-modulated arm, the device is resistance (R)-limited, which could be further optimized by Si selective doping and bringing the Si contact closer to the active region. Hall effect measurements revealed the carrier concentration of the as deposited ITO film, $N_c$ = 2.29 × 10$^{20}$ cm$^{-3}$. The change in the carrier concentration level arising from active capacitive gating is calculated as $\Delta N_c$ = 1.1 × 10$^{20}$ cm$^{-3}$ utilizing both accumulation and depletion mechanisms depending on applied bias. The dynamic switching energy is about 11 pJ/bit using capacitive charging.

The limited ER found indicates that the loss balancing in both of the arms are imperfect, which can be attributed as combinations of several factors such as the passive waveguides being non-identical (sidewalls, roughness, etc.), inadequate Y-splitters skewed astray from 50:50 ratio, fabrication conditions and imperfections, dissimilarity between the as deposited materials (eg. ITO, Au, Al$_2$O$_3$) from used values in the FEM analysis or analytical expressions in design, film quality, non-uniformity of the oxide or metals, etc. The loss imbalance can be estimated form the visibility of the interferometric output. For imbalanced lossy MZ schemes the visibility can be written as [42]

$$v = \frac{1}{\cosh(\Delta\alpha_{bal} L_{bal})} \tag{5}$$

where, $\Delta\alpha_{bal}$ is the amount of loss (absorption) required to bring the system to balance and hence improve the ER, and $L_{bal}$ is the corresponding length needed of the imposed lossy material. We calculate the visibility of the fringes from our results as $v$ = 0.228, which leads to an imbalance factor $\Delta\alpha_{bal} L_{bal}$ of 2.16. This translates to ~2.4 μm of additional Au placed on the active arm to balance the loss and improve *ER*, provided deposited metals follow their exact optical constants from the literature (Fig.

4a). However, this would further limit the output power by enhancing the insertion loss (IL) possibly refraining further detectable measurement.

Possible mechanisms to enhance device performances can include designing the input Y-splitter with a power splitting ratio to compensate the loss resulting as a byproduct of active modulation (Fig. 4b); i.e. evaluate an input Y-splitter with arbitrary splitting ratio, $x$ in terms of the power flow into the active arm. Results show that an input Y-splitter with 75% power flowing into the active arm can compensate for the modulation loss arising from K-K relations towards maximizing *ER*. Furthermore, this approach is beneficial in terms of the IL as it does not have to encounter any parasitic metallic loss as in the present design. Biasing the MOS-stack can propose a challenge in this configuration which can be addressed by a small protrusion of the Si waveguide underneath the active region with selective doping treatments to form a contact pad. Back reflection caused by such protrusion needs to be taken into account also which is not focused on here and can stimulate future research.

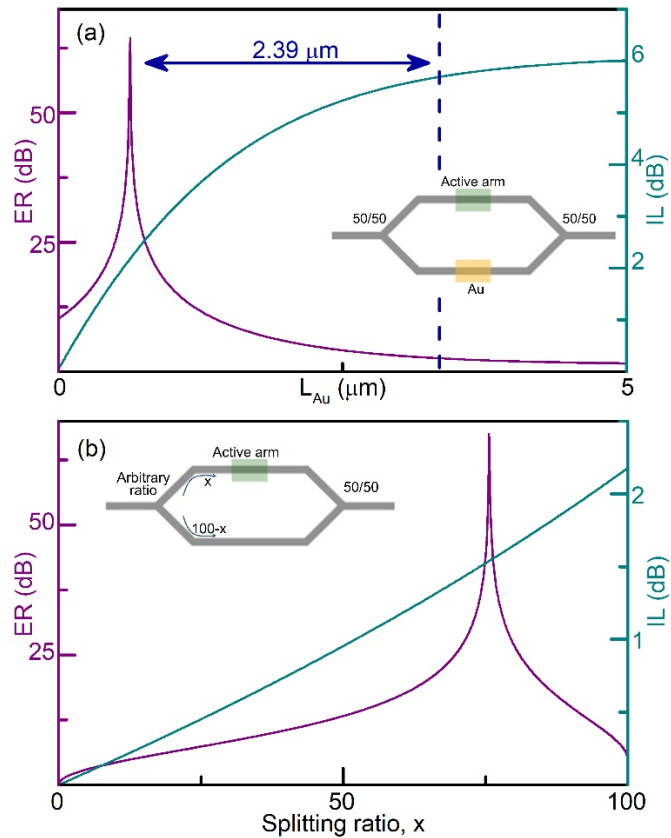

FIG. 4 Extinction ratio, *ER* and insertion loss, *IL* performance for two different loss-balancing methods. (a) The loss is balanced by depositing Au on the passive arm to maximize the ER, however, due to the ohmic loss nature of metal the *IL* is higher compare to second method. The blue dashed line indicates the operating region for our device and additional 2.4 μm of Au can be deposited on the active arm to increase *ER* whereas *IL* would also increase simultaneously. (b) The first Y-branch has arbitrary ratio to maximize *ER*, where the *IL* is also minimized since no extra loss induced on the passive arm. The splitting ratio denotes the power distribution on the active arm in x percentage.

An increase in loss (i.e. decrease in output power) of about 6 dB from the passive to active device was found which is an upper bound IL for the active device. However, the actual IL of the device might be lower as this increase of loss refers to the change in loss from the unprocessed passive MZ structure to the active device after all processing and subsequent processing on the same chip is known to introduce additional loss with every process step (eg. Patterning, liftoff etc.).

**V. CONCLUSION**

In this work, we have demonstrated the first ITO-based Mach Zehnder modulator on a Silicon photonics platform which can be CMOS compatible. Our results confirm a near unity order index change in the ITO material upon applied bias and obtained a low $V_\pi L$ = 0.52 V·mm. Although the speed limitation does not warrant data communication, such response rates (~ms) can avail applications in phased array systems which can be beneficial in emerging technology such as light detection and ranging (LiDAR) for terrestrial and areal localization and mapping.

**METHODS**

**ITO ion beam deposition.** The 10 nm film of ITO was deposited on top of one of the passive waveguides (one arm of the MZ) on SOI substrate at room temperature via ion beam deposition (IBD) using the 4Wave IBD/BTD cluster sputter deposition system. An RF ion gun focused Ar ions onto substrate targets of ITO. The ITO target stoichiometry is 90 wt % $In_2O_3$ / 10 wt % $SnO_2$. A small flow of $O_2$ (2 sccm) was used. The process used an Ar flow rate of 16 sccm, beam voltage of 600 V, beam current of 220 mA and acceleration voltage of 150 V. The sample was set at an angle of 115° and rotated at 10 rpm to ensure smooth profile. The deposition rate used was 0.77 Å/s. The temperature for the process was 20°C to refrain annealing effects. The base vacuum used was $2\times10^{-8}$ Torr and the deposition uniformity was confirmed as 1.5% (1σ) over a 190 mm diameter.

**Fabrication processes.** The pattern transfers were done in e-beam lithography (EBL) using the Raith VOYAGER tool with PMMA based photoresists and MIBK1:3IPA developer for 60 sec. 50 nm of Au for contacts was deposited using an e-beam evaporation system (CHA Criterion). A 3 nm Ti layer was used for adhesion purposes. The $Al_2O_3$ oxide was deposited using the atomic layer deposition (ALD) technique as it provides reliable performance characteristics. The Fiji G2 ALD tool was used at low temperature settings (100°C) for 100 cycles to deposit about 10 nm of $Al_2O_3$ to ensure higher film quality and to avoid any annealing effects to the ITO. Filmetrics F20-UV system was used to characterize $Al_2O_3$ deposition rate. Subsequent etch steps were utilized to etch the oxide film on top of the contacts. We employed a wet etch method using MF319 containing tetramethylammonium hydroxide (TMAH) which reacts with the Al and can etch the oxide thereof, but this is a rather slow process.

**Spectroscopic ellipsometry.** The J. A. Woollam M-2000 DI spectroscopic ellipsometer was used to characterize the optical constants of the deposited ITO thin films as it can provide fast and accurate thin film characterization over a wide spectroscopic range. Variable angle spectroscopic elliposometric (VASE) data was collected from three different angles of 65°, 70° and 75°. A B-spline model was utilized to fit the VASE data and relevant optical constants were found. A mean square error (MSE) value of 1.562 and uniqueness of the fitted thickness parameter coupled with thickness matching from deposition reinforces the fit.

**Resistivity, contact and sheet resistance of ITO films.** Transmission line measurements (TLM) method was used to find the contact resistance of the contact on the ITO pad. A 10 nm thin film was deposited on a $SiO_2$ substrate using the same fabrication processes and contact pads of the same dimensions as present in the device were written using e-beam lithography and subsequent Au deposition and liftoff was also done similar to the actual device. ITO resistivity and sheet resistance was found using the 4-point probe system.

## SUPPLEMENTARY MATERIAL

See supplementary material for additional information.

## ACKNOWLEDGMENTS

Air Force Office of Scientific Research (AFOSR) (FA9550-17-1-0377, FA9550-17-P-0014); Army Research Office (ARO) (W911NF-16-2-0194).

## REFERENCES

[1] R. Amin, C. Suer, Z. Ma, I. Sarpkaya, J. B. Khurgin, R. Agarwal, and V. J. Sorger, "Active material, optical mode and cavity impact on nanoscale electro-optic modulation performance", Nanophot. 7(2), 455-472 (2017).

[2] R. Amin, C. Suer, Z. Ma, I. Sarpkaya, J. B. Khurgin, R. Agarwal, and V. J. Sorger, "A deterministic guide for material and mode dependence of on-chip electro-optic modulator performance", Solid-State Elec. 136, 92-101 (2017).

[3] R. Amin, J. B. Khurgin, and V. J. Sorger, "Waveguide-based electro-absorption modulator performance: comparative analysis," Opt. Exp. **26**(12), 15445-15470 (2018).

[4] L. Zehnder, "Ein neuer interferenzrefraktor," Zeitschrift für Instrumentenkunde **11**, 275-285 (1891).

[5] L. Mach, "Ueber einen interferenzrefraktor," Zeitschrift für Instrumentenkunde **12**, 89-93 (1892).

[6] C. T. Phare, Y.-H. D. Lee, J. Cardenas, and M. Lipson, "Graphene electro-optic modulator with 30 GHz bandwidth," Nat. Phot. **9**(8), 511–514 (2015).

[7] C. Haffner, D. Chelladurai, Y. Fedoryshyn, A. Josten, B. Baeuerle, W. Heni, T. Watanabe, T. Cui, B. Cheng, S. Saha, D. L. Elder, L. R. Dalton, A. Boltasseva, V. M. Shalaev, N. Kinsey, and J. Leuthold, "Low-loss plasmon-assisted electro-optic modulator," Nature **556**, 483–486 (2018).

[8] T. Baba, S. Akiyama, M. Imai, N. Hirayama, H. Takahashi, Y. Noguchi, T. Horikawa, and T. Usuki, "50-Gb/s ring-resonator-based silicon modulator," Opt. Exp. **21**(10), 11869-11876 (2013).

[9] V. J. Sorger, N. D. Lanzillotti-Kimura, R. M. Ma, and X. Zhang, "Ultra-compact silicon nanophotonic modulator with broadband response", Nanophot. 1(1), 17-22 (2012).

[10] Z. Ma, Z. Li, K. Liu, C. Ye, and V. J. Sorger, "Indium-tin-oxide for high-performance electro-optic modulation", Nanophot. 4(1), 198-213 (2015).

[11] R. Amin, M. H. Tahersima, Z. Ma, C. Suer, K. Liu, H. Dalir, and V. J. Sorger, "Low-loss tunable 1D ITO-slot photonic crystal nanobeam cavity", J. Opt. 20(5), 054003 (2018).

[12] C. Ye, K. Liu, R. Soref, and V. J. Sorger, "A compact plasmonic MOS-based 2×2 electro-optic switch", Nanophot. 4(3), 261-268 (2015).

[13] R. Amin, Z. Ma, R. Maiti, S. Khan, J. B. Khurgin, H. Dalir, and V. J. Sorger, "Attojoule-efficient graphene optical modulators," Appl. Opt. 57(18), D130-D140 (2018).

[14] V. J. Sorger, R. Amin, J. B. Khurgin, Z. Ma, H. Dalir and S. Khan, "Scaling vectors of attoJoule per bit modulators", J. Opt. 20(1), 014012 (2017).

[15] E. Feigenbaum, K. Diest, and H. A. Atwater, "Unity-order index change in transparent conducting oxides at visible frequencies," Nano Lett. **10**(6), 2111-2116 (2010).

[16] K. Aoki, J. Kondo, A. Kondo, T. Ejiri, T. Mori, Y. Mizuno, M. Imaeda, O. Mitomi, and M. Minakata, "Single-drive X-cut thin-sheet $LiNbO_3$ optical modulator with chirp adjusted using asymmetric CPW electrode," J. Lightwave Tech. **24**(5), 2233–2237 (2006).

[17] P. Rabiei and W. H. Steier, "Lithium niobate ridge waveguides and modulators fabricated using smart guide," Appl. Phys. Lett. **86**(16), 161115 (2005).

[18] F. Lucchi, D. Janner, M. Belmonte, S. Balsamo, M. Villa, S. Giurgiola, P. Vergani, and V. Pruneri, "Very low voltage single drive domain inverted $LiNbO_3$ integrated electro-optic modulator," Opt. Express **15**(17), 10739–10743 (2007).

[19] N. Anwar, S. S. A. Obayya, S. Haxha, C. Themistos, B. M. A. Rahman, and K. T. V. Grattan, "The effect of fabrication parameters on a ridge Mach-Zehnder interferometric (MZI) modulator," J. Lightwave Technol. **20**(5), 854–861 (2002).

[20] A. Liu, R. Jones, L. Liao, D. Samara-Rubio, D. Rubin, O. Cohen, R. Nicolaescu, and M. Paniccia, "A high-speed silicon optical modulator based on a metal–oxide–semiconductor capacitor," Nat. **427**, 615-618 (2004).

[21] D. Janner, D. Tulli, M. Garcia-Granda, M. Belmonte, and V. Pruneri, "Micro-structured integrated electro-optic $LiNbO_3$ modulators," Laser & Photon. Rev. **3**(3), 301–313 (2009).

[22] X. Xiao, H. Xu, X. Li, Z. Li, T. Chu, Y. Yu, and J. Yu, "High-speed, low-loss silicon Mach–Zehnder modulators with doping optimization," Opt. Express **21**(4), 4116-4125 (2013).

[23] S. Akiyama, T. Baba, M. Imai, T. Akagawa, M. Noguchi, E. Saito, Y. Noguchi, N. Hirayama, T. Horikawa, and T. Usuki, "50-Gbit/s silicon modulator using 250-μm-Long phase shifter based on forward-biased pin diodes," in Proceedings of 9th IEEE International Conference on Group IV Photonics, 192–194 (2012).

[24] L. Alloatti, D. Korn, R. Palmer, D. Hillerkuss, J. Li, A. Barklund, R. Dinu, J. Wieland, M. Fournier, J. Fedeli, H. Yu, W. Bogaerts, P. Dumon, R. Baets, C. Koos, W. Freude, and J. Leuthold, "42.7 Gbit/s electro-optic modulator in silicon technology," Opt. Exp. **19**(12), 11841-11851 (2011).

[25] A. Brimont, D. J. Thomson, F. Y. Gardes, J. M. Fedeli, G. T. Reed, J. Martí, and P. Sanchis, "High-contrast 40 Gb/s operation of a 500 μm long silicon carrier-depletion slow wave modulator," Opt. Lett. **37**(17), 3504–3506 (2012).

[26] S. Dogru, and N. Dagli, "0.77-V drive voltage electro-optic modulator with bandwidth exceeding 67 GHz," Opt. Lett. **39**(20), 6074-6077 (2014).

[27] R. Palmer, L. Alloatti, D. Korn, P. C. Schindler, M. Baier, J. Bolten, T. Wahlbrink, M. Waldow, R. Dinu, W. Freude, C. Koos, and J. Leuthold, "Low power Mach–Zehnder modulator in silicon-organic hybrid technology," IEEE Phot. Tech. Lett. **25**(13), 1226-1229 (2013).

[28] J. Shin, Y.-C. Chang, and N. Dagli, "0.3V drive voltage GaAs∕AlGaAs substrate removed Mach–Zehnder intensity modulators," Appl. Phys. **92**, 201103 (2008).

[29] H.-W. Chen, Y. Kuo, and J. E. Bowers, "Hybrid silicon modulators," Chin. Opt. Lett. **7**(4), 280-285 (2009).

[30] S. Dogru, and N. Dagli, "0.2 V drive voltage substrate removed electro-optic Mach–Zehnder modulators with MQW cores at 1.55 μm," J. Lightwave Technol. **32**(3), 435-439 (2014).

[31] W. M. J. Green, M. J. Rooks, L. Sekaric, and Y. A. Vlasov, "Ultra-compact, low RF power, 10 Gb/s silicon Mach-Zehnder modulator," Opt. Exp. **15**(25), 17106-17113 (2007).

[32] F. Li, M. Xu, X. Hu, J. Wu, T. Wang, and Y. Su, "Monolithic silicon-based 16-QAM modulator using two plasmonic phase shifters," Opt. Comm. **286**, 166-170 (2013).

[33] J. Leuthold, C. Koos, W. Freude, L. Alloatti, R. Palmer, D. Korn, J. Pfeifle, M. Lauermann, R. Dinu, S. Wehrli, M. Jazbinsek, P. Gunter, M. Waldow, T. Wahlbrink, J. Bolten, H. Kurz, M. Fournier, J.-M. Fedeli, H. Yu, and W. Bogaerts, "Silicon-organic hybrid electro-optical devices," IEEE J. Sel. Top. Quan. Elec. 19(6), 3401413 (2013).

[34] C. Haffner, W. Heni, Y. Fedoryshyn, J. Niegemann, A. Melikyan, D. L. Elder, B. Baeuerle, Y. Salamin, A. Josten, U. Koch, C. Hoessbacher, F. Ducry, L. Juchli, A. Emboras, D. Hillerkuss, M. Kohl, L. R. Dalton, C. Hafner, and J. Leuthold, "All-plasmonic Mach–Zehnder modulator enabling optical high-speed communication at the microscale," Nat. Phot. **9**, 525-528 (2015).

[35] H. Kim, and A. H. Gnauck, "Chirp characteristics of dual-drive Mach-Zehnder modulator with a finite DC extinction ratio," IEEE Phot. Tech. Lett. **14**(3), 298-300 (2002).

[36] Y. Zhang, S. Yang, A. E. J. Lim, G. Q. Lo, C. Galland, T. Baehr-Jones, and M. Hochberg, "A compact and low loss Y-junction for submicron silicon waveguide," Opt. Exp. **21**(1), 1310-1316 (2013).

[37] D. J. Thomson, Y. Hu, G. T. Reed, and J. M. Fedeli, "Low loss MMI couplers for high performance MZI modulators," IEEE Phot. Tech. Lett. **22**(20), 1485-1487 (2010).

[38] J. A. Dionne, K. Diest, L. A. Sweatlock, and H. A. Atwater, "PlasMOStor: a metal−oxide−Si field effect plasmonic modulator," Nano Lett. **9**(2), 897–902 (2009).

[39] S.-K. Koh, Y. Han, J. H. Lee, U.-J. Yeo, and J.-S. Cho, "Material properties and growth control of undoped and Sn-doped $In_2O_3$ thin films prepared by using ion beam technologies," Thin Solid Films **496**(1), 81-88 (2006).

[40] S. H. Lee, S. H. Cho, H. J. Kim, S. H. Kim, S. G. Lee, K. H. Song, and P. K. Song, "Properties of ITO (indium tin oxide) film deposited by ion-beam-assisted sputter," Mol. Cryst. Liq. Cryst. **564**(1), 185-190, (2012).

[41] S. L. Chuang, Physics of optoelectronic devices (Wiley, New York, 1995).

[42] P. Berini, "Bulk and surface sensitivities of surface plasmon waveguides," New J. Phys. **10**, 105010 (2008).